\documentclass[12pt,preprint]{aastex}

\newcommand{\Ni}{\rm ^{56}Ni}
\newcommand{\co}{\rm ^{56}Co}
\newcommand{\fe}{\rm ^{56}Fe}
\newcommand{\nni}{{\rm Ni}}

\newcommand{\carbon}{{\rm C}}
\newcommand{\oxygen}{{\rm O}}
\newcommand{\tni}{t_{Ni}}

\newcommand{\tco}{t_{Co}}
\newcommand{\qni}{Q_{Ni}}
\newcommand{\qco}{Q_{Co}}

\newcommand{\Msun}{~M_\odot}
\newcommand{\msun}{M_\odot}
\newcommand{\gcm}{\rm ~g~cm^{-3}}
\newcommand{\dyc}{\rm ~dyn~cm^{-2}}

\newcommand{\kms}{\rm ~km~s^{-1}}
\newcommand{\ergs}{\rm ~erg~s^{-1}}
\newcommand{\ergcc}{\rm ~erg~cm^{-3}}

\begin{document}


\title {CLUMP DEVELOPMENT
 BY THE NICKEL BUBBLE EFFECT IN SUPERNOVAE \\
to be published on ApJ 10 June 2005, v 626 1 issue}

\author{Chih-Yueh Wang}
\affil{Department of Physics \\
Chung-Yuan Christian University \\
Chungli, Taiwan 320}

\begin{abstract}

We used one-dimensional radiative-transport
radiation hydrodynamical simulations to investigate
the formation of clumping in freely-expanding supernova ejecta due to 
the radioactive heating from the $\Ni \rightarrow\ \co \rightarrow\
\fe$ decay sequence.
The heating gives rise to an inflated Nickel bubble, which induces 
a  forward shock that 
compresses the outer ambient gas into a  shell.
The radiative energy deposited by the radioactivity
leaks out across the shock by radiative diffusion, and 
we investigate its effect on the evolution of the ejecta structure. 
Compared to the hydrodynamical adiabatic approximation with $\gamma =4/3$, 
the preshock gas becomes accelerated by the radiation outflow.
The shock is thus weakened and the shell becomes broader and less dense. 
The thickness of the shell takes up 
 $ \la 4\%$ of the radius of the bubble, 
and the structure of the shell can be approximately described
by a self-similar solution.
We compared the properties of the shell components
with those of the ejecta clumps indicated by our previous 
 hydrodynamical simulations for the interaction of clumps
 with the outer supernova remnant.
 The high density contrast across the shell, $\chi \sim 100$, 
 is compatible with that of ejecta clumps as indicated for Tycho's knots, 
  but there is insufficient dense gas to cause a pronounced 
  protrusion on the outline of a core collapse supernova remnant, 
  like the bullets in the Vela remnant.

\end{abstract}


\section{INTRODUCTION}

Observations of SN 1987A showed that the distribution of Fe in the ejecta 
is not what
would be expected in the simplest models: it extended to higher velocity
than expected and had a large filling factor for its mass of
$0.07\Msun$ determined from the supernova light curve (McCray 1993;
Li, McCray, \& Sunyaev 1993).
A plausible mechanism for the large filling factor is the Ni bubble
effect, in which the radioactive progenitors of the Fe expand
relative to their surroundings because of the radioactive
power deposition (Woosley 1988; Li et al. 1993; Basko 1994).
This effect is important during the first $\sim 10$ days after the supernova,
when the radioactive power is significant and  the
 diffusion of energy is not yet an important process.

An expected effect of the Ni bubble expansion is to create clumps
in the nonradioactive gas.
There is widespread evidence that the ejecta of core collapse supernovae
are clumpy.
The oxygen line profiles in the nearby Type II supernovae SN 1987A
(Stathakis et al. 1991) and SN 1993J (Spyromilio 1994;
Matheson et al. 2000) showed 
evidence for structure, implying
that the gas is clumped.
The velocity range for the emission extends to 
$1,500\kms$ in SN 1987A and $4,000\kms$
in SN 1993J.
Similar evidence for clumping has  been found in the Type Ib
supernova SN 1985F (Filippenko \& Sargent 1989).
Among young supernova remnants, Cas A is the prototype of the
oxygen-rich SNRs, which show evidence for freely expanding, oxygen-rich
ejecta in clumps.
Puppis A, with an age $\sim 3,700$ yr, is a more elderly example of such
a remnant (Winkler et al. 1988).

Wang \& Chevalier (2001, 2002, hereafter WC01, WC02) 
investigated the role of clumps in the evolution of supernova remnants.
In Tycho's remnant, believed to be the remnant of a Type Ia supernova, the
presence of ejecta knots near the outer shock front 
 requires a   density contrast $\chi\ga 100$ relative to the surrounding ejecta (WC01).
 The remarkable protrusions in the X-ray image of the Vela remnant are likely
 to be caused by ejecta clumps (Aschenbach et al. 1995) and WC02
 found that 
$\chi \sim 1000$ is needed to create the structures. 
The free expansion velocities for the clumps were estimated at
$\sim 3,000\kms$.

Our aim here is to investigate whether the Ni bubble effect can
cause the clump structure apparently needed in the Tycho and Vela remnants.
We compute 
 one-dimensional hydrodynamical calculations, building on the 
 work of Basko (1994), suitable for core collapse
supernovae.                                         
We extend Basko's work by including the effects of radiative
diffusion and by carefully examining the density structure
of the shell swept up by the Ni bubble.
We further consider Type Ia supernovae using the same
methodology. 
We show our computational setup and methods in \S~2.
The solutions for radiation-hydrodynamical effects are given in \S~3.
 We also draw on the analogy to power input in a pulsar bubble to provide insight into the shell structure and evolution.
In \S~4 and \S~5, we discuss the radiative effects on the
Rayleigh-Taylor instabilities of the Ni bubble shell and
the inferred ejecta-clump properties.
Our conclusions are in \S~6.

\section {INITIAL CONDITIONS AND METHODS}
 
\begin{figure}[!hbtp]
\centerline{\includegraphics[width=10cm]{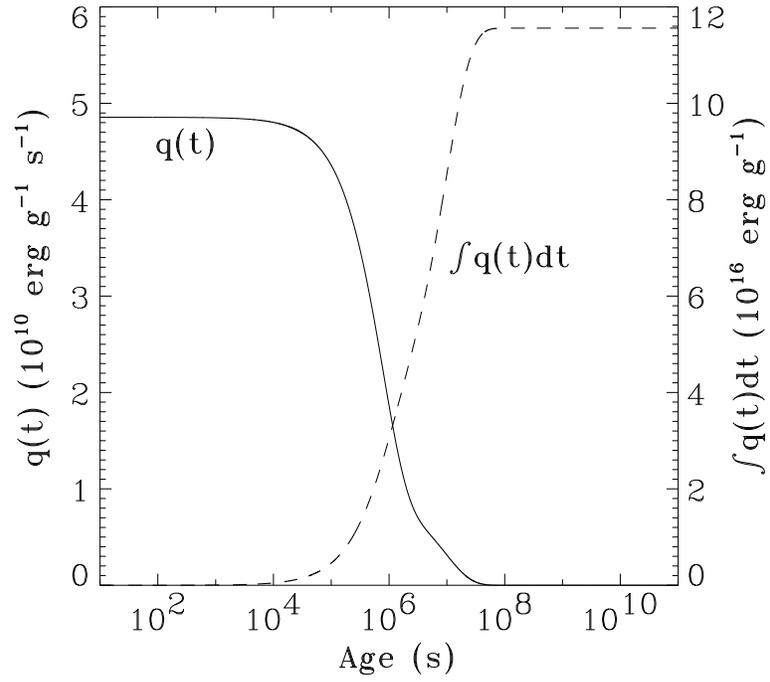}} 
\caption{\small
Evolution of the instantaneous and accumulated heating rate of the Nickel and
Cobalt radioactive decay.
\label{fig.bradio}}
\end{figure}

We consider all of the $\Ni$ synthesized in a
supernova explosion initially resides in 
an isolated sphere
in the center of supernova ejecta, 
with a mass $M_{Ni}$ and a density
contrast $\omega$ relative to the surrounding 
diffuse ejecta. 
The total radioactive energy released in the 
$\Ni \rightarrow\ \co \rightarrow\ \fe$ decay sequence
is deposited at a rate $q(t)$ (Basko 1994)
\begin{equation}
q(t)= {\qni \over \tni} \exp\left( -{{t\over \tni}} \right) + {\qco \over \tco}
      { \exp(-t/\tco)-\exp(-t/\tni) \over 1-\tni/\tco },
\end{equation}
where 
$\tni$,$\tco$ and $\qni$,$\qco$ are, respectively, the mean life and 
specific decay energy
of $\Ni$,$\co$:
$$
       \left\{
        \begin{array}
                     {l@{\quad\quad}l}
        \ \tni  = 7.6\times 10^{5} \rm \sec \ = \ 8.8 \ \rm days^\dagger 
        \qquad &  \Ni \rightarrow\ \co \\
     \qni   = 2.14 {\rm \ MeV/decay} = 3.69\times 10^{16} \rm \ ergs \ g^{-1}   
        \end{array}   \right\}
$$
$$
       \left\{
        \begin{array}
                     {l@{\quad\quad}l}
        \ \tco = 9.64 \times 10^{6} \rm \sec \ = \ 111.5 \ \rm days  
         \qquad  &  \co \rightarrow\ \fe \\
     \qco = 4.57 {\rm \ Mev/decay}   = 7.87\times 10^{16} \rm \ ergs \ g^{-1}.
                     \end{array} \right\}
$$
  (Lide 1992; $\dagger$ 8.5 days, Fireston \& Shirley 1996). 
  The accumulation of the total radioactive energy rises from $10^4$ sec and 
  saturates at $\ga 10^7$ sec 
 (about 25\% of the total radiative energy is deposited
  between $10^7$ and $10^8$ sec, Fig.~\ref{fig.bradio}).
Before the saturation point is reached, the radioactive energy  can
leak out of the bubble by radiative diffusion.
This can occur while the gas remains optically thick
to the $\sim$ 1 MeV $\gamma$-rays.

We assume that the diffuse supernova ejecta are initially freely-expanding  
so that each gas element moves with a constant velocity $v=r/t$.
The part of the ejecta unaffected by the  radioactivity 
 has its density
drop as $t^{-3}$, and the pressure drops as $t^{-4}$ for a $\gamma=4/3$, 
radiation
dominated fluid. 
We first refer our simulation initial conditions to Basko's parameters  
for SN 1987A (Basko 1994 and the references therein): 
$$ 
M_{Ni}=0.075\Msun, \qquad \omega=3 
$$
$$
t_{0}=10^{4} {\rm \ sec}, \qquad \rho_{a0}=10^{-4} \gcm,  \qquad {\it P_{0}}=2.5 \times 10^{11} {\rm \ dyn \ cm^{-2}},
$$ 
where 
$t_{0}$, $\rho_{a0}$ and $P_{0}$ are the initial age, 
ejecta substrate density (surrounding the bubble), and
background pressure, respectively.  
The initial velocity at the bubble edge $r=R_{0}$ is 
given by the initial $\Ni$ density contrast and age:
$$
U_{0} = {R_{0} \over t_{0}} = 
\left(3M_{Ni}\over 4\pi \rho_{Ni}\right)^{1/3}{1\over t_0}
=7 \times 10^{7} \ \omega^{-1/3} \rm \ cm \ s^{-1}.
$$
For $\omega=3$, 
$R_0=5.0 \times 10^{11} \rm \ cm$.
The initial pressure is assumed
uniformly distributed inside and outside the bubble,
independent of the bubble size.
The pressure in the Ni bubble is determined by the radioactive
power input at later times.

We note that the above parameters give an ejecta structure comparable 
to the inner flat-density component of the power-law ejecta density model
that has a power index $n=8$ ($\rho\propto r^{-n}$ in
the outer parts), an ejecta mass $M=10 \Msun$,
and an explosion energy $E=10^{51}$ ergs (Chevalier \& Liang 1989).
The transition between the flat and power law parts of the density
profile occurs at $v = 3162 \kms$ for these parameters.
This set of explosion conditions is suitable for
core collapse supernovae and
had been used in our two-dimensional hydrodynamical 
simulations for the Vela supernova remnant (WC02). 
The energy density is initially dominated by the energy deposited by the supernova shock.
Shigeyama \& Nomoto's (1990, their Fig. 2) model for SN 1987A 
indicates an 
adiabat 
$\kappa = 6.3 \times 10^{15}$ (cgs units) where $P = \kappa \rho^\gamma$;
 for $t_0=10^4$ sec, 
$\rho_{0} = 9.4\times 10^{-5}\gcm$ in our density model, and so
$P_{0} = 2.7\times 10^{10}$ dynes cm$^{-2}$. 
We take $P_{0} = 2.7\times 10^{10}$ dynes cm$^{-2}$ as the appropriate value at 
$10^{4}$ sec.
For $t_0 = 100$ sec, $P_{0} = 2.7 \times 10^{18}$ dynes cm$^{-2}$.             
The time to set up the free expansion phase depends on the radius of the
star because of the effects of reverse shock waves.
For a blue supergiant progenitor star, like that of SN 1987A, the timescale
is $\sim 10^4$ sec; for a red supergiant progenitor, the timescale
is $\sim 10^5$ sec.
It is needed to initiate the simulations at $t \sim 10^2$ sec
to acquire a numerically saturated solution at 
$t \ga 10^4$ sec (the age of interest). 
The use of a freely-expanding
velocity prior to this stage should not affect the 1-D symmetrical solution,
considering that the radioactive heating is strong.
Models of
 supernova explosions predict smooth
 56Ni profiles in both thermonuclear and core collapse supernovae, 
 with the density contrast $\omega$ below a factor of 3 
 (Nomoto, Thielemann \& Yokoi 1984; Woosley 1988).
  We vary $\omega$ 
  in the range of $0.1 \le \omega \le 3$, 
 with the other initial parameters scaled accordingly 
 to the standard values described above.


Next we consider the initial parameters for Type Ia supernovae. 
We assume a powerlaw density profile $n=8$ as
in core collapse supernovae,                       
with $M=1.4\Msun$, $E=10^{51}$ ergs, and $M_{Ni}=0.5\Msun$. 
The transition between the flat and power law parts of the density
profile occurs at $v = 8452 \kms$ for these parameters.                       
The background pressure in Type Ia ejecta is not well documented.
   To estimate, we make use of
   the entropy change $dS=dQ/T$ during the nuclear burning from C/O into Ni.
   During the process, $kT \sim 0.6$ MeV,
   and the energy released per nucleon is
   \begin{equation}
     dQ \sim  \left [2m(\carbon)+2m(\oxygen)-m(\nni) \right] c^{2} / 56 
        \sim 0.79 \ \rm MeV/nucleon, 
   \end{equation}
   where $m(\rm X)$ is the mass of element X;
     $m(\carbon)=12.0000 \ \rm amu$, $m(\oxygen)=15.9949 \ \rm amu$,
     and $m(\nni)=55.9421 \ \rm amu $ 
     (Audi et al. 2003),  
    and $c$ is the speed of light. 
  Then $dS \sim 1.3 k \rm \ /nucleon  = 1.8 \times 10^{-16} \rm erg/K/nucleon$,
  where $k$ is the Boltzmann constant  
  (Shigeyama, private communication).
  The adiabat $\kappa$ was obtained by relating
  the entropy per nucleon
  $S=[(4aT^{3}/3)(m(\nni)/56) / \rho] $, where the initial entropy before 
  the burning is ignored, 
  to the radiative pressure $P=aT^4/3$  
  and $P=\kappa \rho^{4 \over 3}$.
  Subsituting $P$ and $\kappa$ into $S$,
  we find
 \begin{equation}
 \kappa = \left[ \ ({3 \over a})^{1/4}({S \over 4 (m(\nni)/56)}) \right]
           ^{4/3} \sim 6.1 \times 10^{14} \ \rm (cgs \ units).
 \end{equation}
  For $t_0=10^4$ sec,
  $\rho_{0} = 6.92\times 10^{-7}\gcm$ in our density model, and so
           $P_{0} = 3.7 \times 10^{6}$ dynes cm$^{-2}$.
  The adiabat (and so the background pressure) could be higher, considering
  there is an initial entropy distribution that acts to freeze 
  the deflagration front.
  The background energy density has the profile of the mass density because
  of the assumed uniformity.

We first approach the problem using hydrodynamical simulations 
with $\gamma = 4/3$
in both the bubble interior and the ejecta substrate.
We follow the bubble-shell interface on a spherical expanding grid,
and add the specific radioactive energy 
 to the internal
 energy within the bubble uniformly and locally, i.e., in proportion 
to the local mass density.
A reflecting condition is used for the inner boundary, and 
a non-zero gradient outflow condition is invoked on the outer boundary
to eliminate the spurious shock raised by the grid expansion.

We further include radiation hydrodynamics (RHD) in the simulations using 
the  ZEUS2D 
code (Stone, Mihalas, \& Norman 1992).
 The radioactive power is deposited into the radiation
 instead of the material and $\gamma =5/3$ for the matter.
The code is based on finite difference and finite volume on an 
Eulerian grid and 
uses an artificial viscosity to smooth shock transitions.
The RHD algorithm employs 
nonrelativistic full 
radiative transport,
under the assumption of LTE (local thermal equilibrium) and gray opacity,
 and is thus equally applicable to 
 both optically thin and thick media. 
 The flow evolves towards the optically thin regime where
 radiation tends to stream away across the Ni bubble shell.
 The full transport scheme ensures the proper evolution 
 while the radioactive energy is continually added within the bubble.

At the age of interest, the opacity is
dominated by electron scattering.
A maximum absorption 1/1000 times the magnitude of scattering is applied in our calculation. 
For completely ionized heavy elements, the Thompson scattering 
gives an opacity
$0.2 \ \rho(r,t) \rm \ cm^{-1}$, where $\rho(r,t)$ is the material density. 
The initial energy is equally partitioned 
between radiation and material.   
Considering that a Type Ia has lower temperature and smaller radius, 
$P_{rad} \sim r^{-4}$ and $P_{gas} \sim r^{-5}$, 
the equi-partition assumption may be more valid
for core collapse supernovae. 

 In the RHD algorithm, the coupled radiation and material energy densities
 are solved using Newton-Raphson iteration for convergence.             
   When the radiation pressure is in excess of  
  the gas pressure  
  by several orders of magnitude, 
  convergence becomes extremely difficult to reach.
  We thus stop the simulations prior to the optically thin stage.
In our situation,
the radiation effects are most important when
the gas is still optically thick, so that the diffusion approximation
is adequate. 
The choice of the flux limiters from which the Eddington tensor 
is computed does not affect the diffusion result. 
For details of radiation hydrodynamics, we refer readers to Stone
et al. (1992) and the references therein. 

\section {SHELL STRUCTURE AND EVOLUTION} \label{sec:resu}

\subsection {Simulations} \label{subsec:pow}

\begin{figure}[!hbtp]
\centerline{\includegraphics[width=10cm]{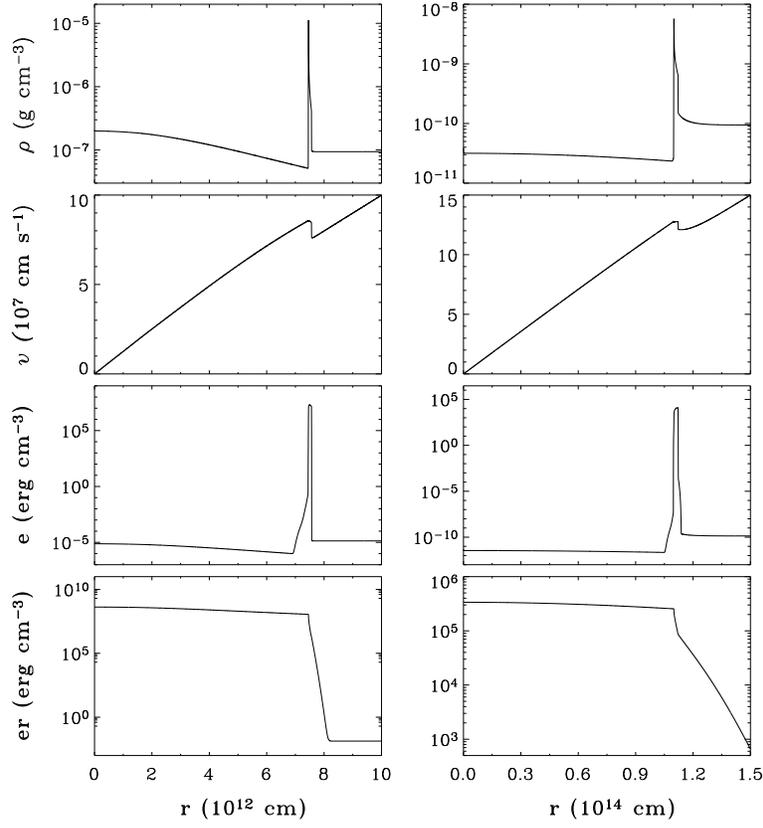}} 
\caption{\small
Distributions of density, velocity, gas energy density, and radiation energy
density of the Nickel bubble structure in the flat-density ejecta of the
core collapse supernova
at $10^{5}$ and $10^{6}$ sec. The initial Nickel mass is
$M_{n}=0.075\Msun$. The grid has 8000 uniform zones that resolve the shell into $\sim 100$ zones.
\label{fig.prof}}
\end{figure}
\begin{figure}[!hbtp]
\centerline{\includegraphics[width=10cm]{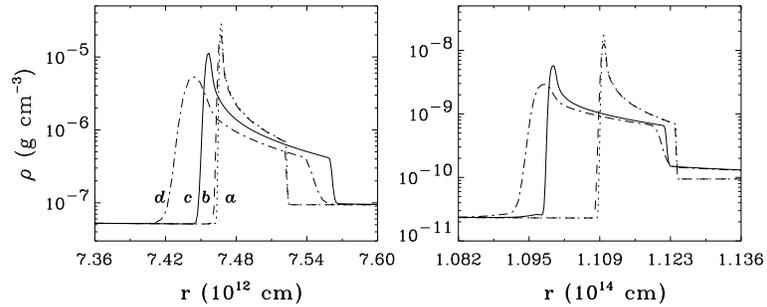}} 
\caption{\small
Density profiles of the Nickel bubble shell in the core collapse SN model
at $10^{5}$ and $10^{6}$ sec.
a: HD solution on a nonuniform grid of 8000 zones that resolves the shell into
200 zones.
b: HD solution on a uniform grid of 8000 zones.
c: RHD solution on a uniform grid of 8000 zones.
d: RHD solution on a uniform grid of 2000 zones.
\label{fig.profdens}}
\end{figure}
\begin{figure}[!hbtp]
\centerline{\includegraphics[width=10cm]{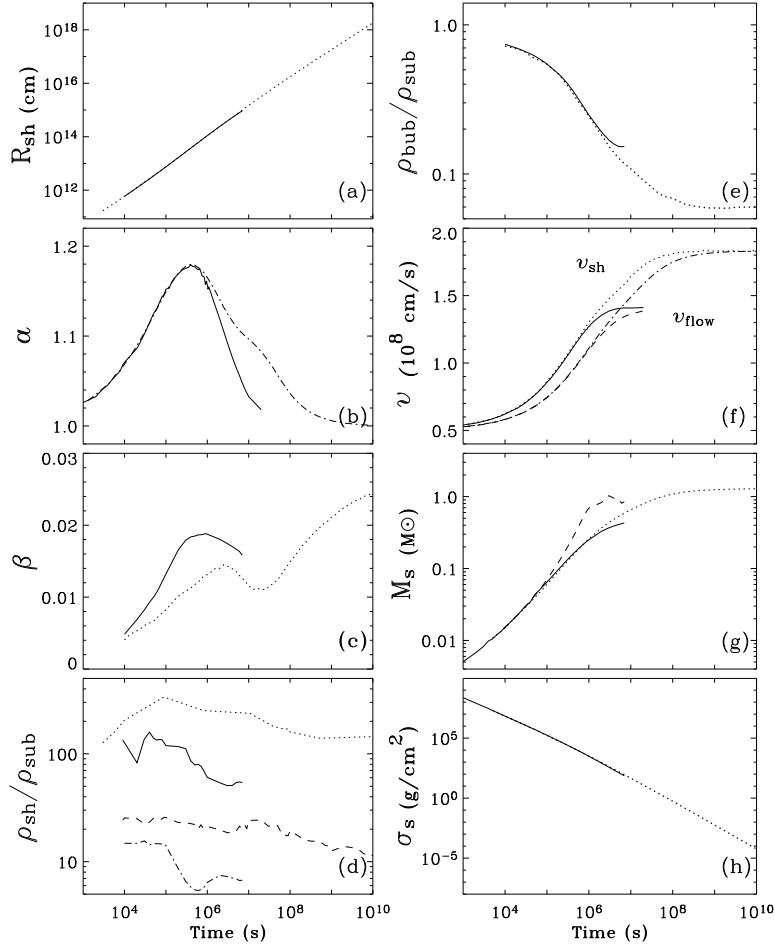}} 
\caption{\small
Evolution of the dynamical properties of the Nickel bubble shell
in the core collapse SN model.
Solid and dotted lines represent the RHD and the HD case, respectively.
(a)
Radius of the shock $R_{sh}$.
(b)
Expansion rate of the shell, $a=dlnR_{sh}/dlnt$,
equivalent to the velocity contrast between the shock
and the ambient free-expanding ejecta.
(c)
Thickness ratio $\beta$
of the shell. The value is influenced by numerical resolution.
(d)
Density contrast
between the densest region in the shell and the ejecta substrate.
Dashed and dash dot
lines represent the average density contrast in the HD and the RHD case,
respectively.
(e)
Density contrast
between the bubble and the ejecta subtrate.
(f)
Shell velocity $v_{sh}=dR_{sh}/dt$ (left)
and flow velocity $v_{flow}= R_{sh}/t $ (right)
at the shock front.
(g)
Swept-up mass $M_s$ in the shell.
In the RHD case, two estimates are given for $M_s$
due to the lower resolution.
Solid line: estimated within the inner sharp edge of the shock front.
Dashed line: estimated within the outer edge of the shock
front where the density flattens out.
(h)
Surface density $\sigma_s \equiv M_s / 4 \pi R_{sh}^2$ of the shell.
\label{fig.big4}}
\end{figure}

We show  in 
Figs.~\ref{fig.prof}, \ref{fig.profdens}, and \ref{fig.big4} 
the ejecta structure and evolution properties
in the core collapse supernova model using 
$n=8$, $E=10^{51} \ergs$, and $M=10\Msun$.
These simulations were initiated with
$t_0=10^2$ sec, 
 $M_{Ni}=0.075\Msun$ and $\omega=3$.
The diffuse ejecta have an initial density $\rho_0=94.4 \gcm$ and 
an accumulated radioactive energy 
density $e_{rad0} = 1.38 \times 10^{15} \ergcc$.
The initial material and radiation energy density in the background 
are equally divided,
$e_0$ = $e_{r0}$ =$1.35 \times 10^{10} \ergcc$. 

We first examined the evolution of the flow in the purely 
hydrodynamic simulations. 
 The expansion of the bubble  
 gives rise to a strong forward shock 
 behind which the ambient gas is compressed into a dense shell
 (Figs.~\ref{fig.prof} and \ref{fig.profdens}).
The inner edge of the shell is a contact discontinuity where
 the gas comoves with the bubble-shell interface; it has a
 high density because of the acceleration of the shell.
The shock front is continually accelerated outward
(Fig.~\ref{fig.big4}(a)). 
To describe the acceleration, we use
$a\equiv (dR_{sh}/dt)/(R_{sh}/t)$,
 the expansion rate evaluated at the shock front $R_{sh}$, 
 equivalent to 
the velocity ratio of the shell to the preshock diffuse ejecta.
We find that $a$ rises to 
a maximum $\la 1.20$ around
$\la 10^{6}$ sec, and subsequently tends 
to 1.0 at late times as the shell becomes comoving with the preshock gas
(Fig.~\ref{fig.big4}(b)). 
Because the background pressure has little influence on the dynamics,
the shell structure becomes frozen into the flow.  

The shell is very thin, with a thickness as a fraction of the radius
$\beta \equiv h_{sh}/R_{sh} \la 0.02$,
where $R_{sh}$ and $h_{sh}$ are respectively 
the radius and
thickness of the shell (Fig.~\ref{fig.big4}(c)). 
The initial shock compression is set by the jump condition
\begin{equation}
{\rho_1\over\rho_0}={(\gamma +1){\cal M}^2\over
(\gamma -1){\cal M}^2 +2},
\end{equation}
where
\begin{equation}
{\cal M}={v_{sh}\over c_0}=\left({dR_{sh}\over dt} -{R_{sh}\over t}\right)
\left(\rho_0 \over \gamma p_0\right)^{1/2}
\end{equation}
is the Mach number of the shock, $\rho_1$ is the postshock density,
$v_{sh}$ is the shock velocity relative to the preshock gas,
and $c_0$ is the sound speed in the preshock medium.
For a radiation dominated, strong shock, $\rho_1/ \rho_0 = 7$.
The accelerating shell has a higher density in the postshock region
so that
a typical compression factor in the shell (relative to the preshock
gas) is $\chi \sim 10$ (Fig.~\ref{fig.big4}(d)).
The compression factor in the bubble is  $\chi \sim 0.1$
(Fig.~\ref{fig.big4}(e)).
We note that $\beta\approx (3\chi)^{-1}$.
The highest density occurs at the inner edge, where the density is
limited by the numerical resolution (Fig. 3).
 
In the RHD case,
 the  energy from the heated gas
leaks out from the bubble energy reservoir as time progresses (Fig. 2).
The gas is radiation dominated.
Radiative diffusion allows the internal energy to propagate ahead 
of the shock wave, 
and so eliminates the temperature and the radiation pressure gradients.
In order for diffusion to be
dynamically important, 
it requires that the diffusion time scale, $t_d$, be smaller than 
the hydrodynamic time scale of the flow, $t_{h}$;
i.e.,
\begin{equation}
t_d \approx {h_{sh}^{2} \over \lambda c} < t_h \approx {h_{sh} \over v_{sh}} 
\label{diffusiontime}
\end{equation}
where $\lambda$ is the photon mean free path.
We note that
\begin{equation}
{t_d\over t_h}\approx \left(h_{sh}\over \lambda\right)
\left(v_{sh}\over c\right).
\end{equation}
At $t \approx 3 \times 10^{6}$ s,
$h_{sh} = 10^{13}$ cm,
$\lambda = 1/ \kappa \rho = 5 \times 10^{10} $cm,
so that $t_{d} = h_{sh}^{2}/ \lambda c = 6 \times 10^{4}$ s.
At this time, $v_{sh}\approx 1\times 10^7$ cm s$^{-1}$, so
$t_d / t_h \approx 0.07$.
The gas is optically thick, but radiative diffusion is important.
As a result of diffusion, the preshock gas 
is accelerated by the radiation;
the shock is then  weakened.
The shock front develops a radiative precurso (Fig. 2).
The acceleration of the bubble shell is reduced 
because of the radiative loss.
The rise in the preshock sound speed and the acceleration of
gas ahead of the shock reduced the
shock compression (eqn. [4]).
The shell  becomes broader and less dense compared to the HD case 
 (Fig.~\ref{fig.big4}(c)).
The sharp density jump at the shock front in the RHD case is also smoothed out
before the gradient of $e_r$ diminishes to zero (Fig.~\ref{fig.big4}(d)).
The gas tends toward a freely expanding state 
in which the density
profile is frozen into the flow. 

\begin{figure}[!hbtp]
\centerline{\includegraphics[width=10cm]{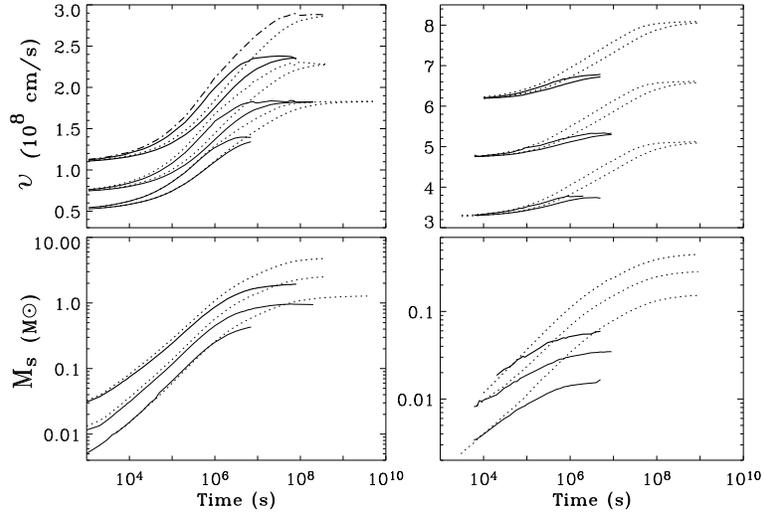}} 
\caption{\small
Evolution of shell and shock velocity and accumulated mass in the
shell with varying initial Ni abundance.
Solid and dotted lines represent the RHD and the HD case, respectively.
Left:
$M_{Ni}$ = 0.075 (bottom line), 0.225 (middle line), and 0.75 $\msun$ (top line)
 for the core collapse SN model.
Right:
$M_{Ni}$ = 0.075 (bottom line), 0.225, (middle line), and 0.5 $\msun$ (top line)
 for the Type Ia SN model.
\label{fig.mni}}
\end{figure}

The diffusion effect speeds up the freeze-out of the ejecta structure.
For our parameters, 
  the freezing occurs 
  at a velocity $v_f \sim 1500\kms$ and a time $t_f \sim 10^{7}$ s, 
significantly 
earlier than
the HD case where
  $t_f \sim 10^{9}$ s and 
 $v_f \sim 1900 \kms$ (Fig.~\ref{fig.big4}(f)).
  In the HD case, the expansion sustains   
  until pressure equilibrium is reached.
Both cases show a density contrast $\chi \ga 10$  
between the shell and the preshock ejecta.
The inner edge (contact discontinuity) of the shell 
is the densest with $\chi \ga 100$ .
The shell thickness is increased 
to $\sim$ 4\% of the radius of the bubble.
The swept-up mass in the shell remains about $M_s\sim 1 \Msun$
(Fig.~\ref{fig.big4}(g)).
Note that we define the thickness
as the distance from the contact discontinuity to 
the postshock outer dense region at the shock front, 
considering that the outermost postshock density profile 
flattens out into the ejecta substrate.
   At $\la 10^7$ sec, the surface density of the shell
   drops below $\sim 17-33g/cm^2$,
   the mean free path for the $\sim 1 MeV$ $\gamma$-ray photons
  (Woosley, Pinto, \& Hartmann 1989; Shigeyama \& Nomoto 1990). 
   We stopped the RHD calculations prior to this time (Fig.~\ref{fig.big4}(h)).
When the initial Nickel mass is increased to
 3 times $0.075\Msun$, the onset of 
 freezing delays to 
$t_f \sim 10^{8}$ s with $v_f \sim 1800 \kms$ and $M_s \la 3 \Msun$.
 For a 10 times increase, $t_f \sim  10^{8}$ s, $v_f \sim 2400 \kms$, and
$M_s \sim 5 \Msun$ (Fig.~\ref{fig.mni}). 
It is therefore expected that the radiation relaxes before $10^{8}$ s, given 
any large Ni overabundance.
   The reason is that 
   the photon mean free path is increased by the broadening of the shell,
   and so diffusion becomes more rapid with time. 

\subsection{Effect of Radiation}                                                                                    
 The effect of radiative diffusion does not
 increase the density contrast
 of the Nickel bubble shell,
 as in shell formation during
 shock break-out in a Type II supernova where radiative energy losses
 by diffusion lead to a large compression that is eventually limited
 by the gas pressure (WC01, Chevalier 1976). 
 In the latter case, the flow behind the shock wave is being
 decelerated, and
 the immediate postshock region is being compressed.   
 A drop in the dominant radiation pressure then is compensated 
 by compression of the gas to the point where gas pressure becomes
important, resulting in 
 a large density enhancement. 
 In our case, the pressure gradient
 has an opposite sign; the pressure decreases outward from the contact 
 discontinuity to the shock front,
and the radiative loss to the preshock gas is replenished by   
 the  energy diffusing from the bubble.
 The diffusion of energy ahead of the shock front leads to a
 weaker shock front and less compression in the shell. 

\subsection{Models for Type Ia supernovae}

 We investigated 
 the ejecta structure for Type Ia SNe 
 using the inner flat component of the power law
 model, $n=8$, $E=10^{51}$ ergs, and $M=1.4 \Msun$. 
  We present the evolution of the dynamical properties 
  in Fig.~\ref{fig.big4-Iapow}.
  The simulations were our highest resolved runs
  initiated with $t_{0}=10^2$, 
  $M_{Ni}=0.5\Msun$, and $\omega=1.5$. The initiated time gives 
  an initial mass density 
  $\rho_0=6.92 \times 10^{-1} \gcm$,
  and an initial material and radiation energy
  density $e_0$ = $e_{r0}$ = $ 5.5 \times 10^{14} \ergcc$ 
  for a standard adiabat $\kappa = 6.1 \times 10^{14} \rm (cgs \ units)$.
 Here a lower background pressure was used, 
  $e_0 = 1.3 \times 10^{13} \ergcc$. 
The initial $\Ni$ has an accumulated radioactive energy
density $e_{rad0} = 3.15 \times 10^{9} \ergcc$.

\begin{figure}[!hbtp]
\centerline{\includegraphics[width=10cm]{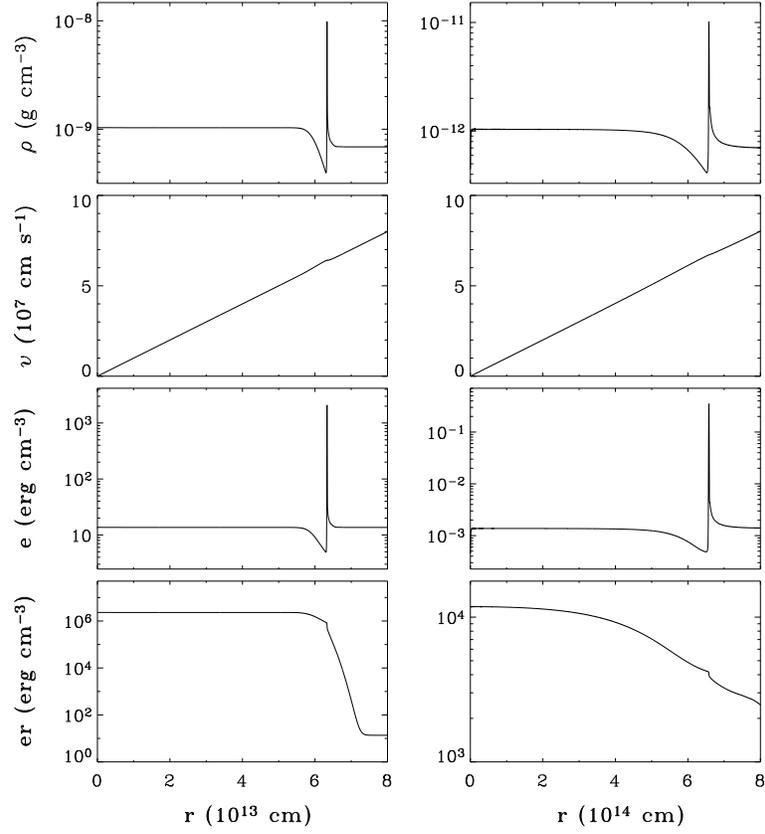}} 
\caption{\small
Distributions of density, velocity, gas energy density, and radiation
energy density of the Nickel bubble structure in
the Type Ia SN model
at $10^{5}$ and $10^{6}$ sec. The initial Nickel mass is
$M_{n}=0.5\Msun$. The grid has 16000 uniform zones.
\label{fig.prof-Iapow}}
\end{figure}

\begin{figure}[!hbtp]
\centerline{\includegraphics[width=10cm]{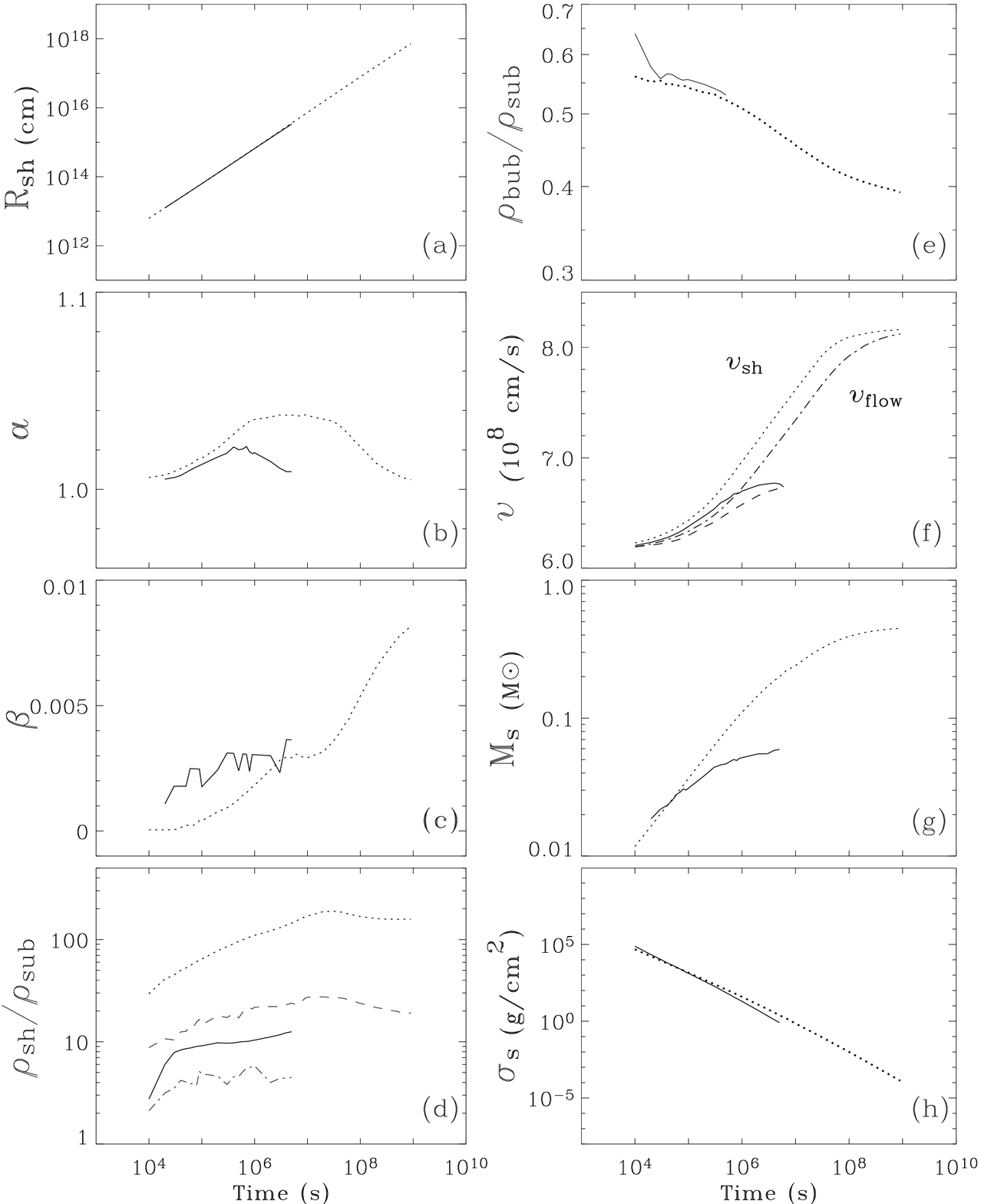}} 
\caption{\small
Evolution of the properties of the Nickel bubble shell in the Type Ia SN model.
Solid and dotted lines represent the RHD and the HD case,
respectively.
(a)
Radius of the shock $R_{sh}$.
(b)
Expansion rate of the shell, $a=dlnR_{sh}/dlnt$,
equivalent to the velocity contrast between the shock
and the ambient free-expanding ejecta.
(c)
Thickness ratio $\beta$
of the shell. The thickness is estimated within the densest point and
the outer
edge of the shock front. In the RHD case it is not well defined due
to a low resolution.
(d)
Density contrast
between the densest region in the shell and the ejecta substrate.
Dashed and dash dot
lines represent the average density contrast in the HD and the RHD case,
respectively.
(e)
Density contrast between the bubble and
the ejecta subtrate.
(f)
Shell velocity $v_{sh}=dR_{sh}/dt$ (left)
and flow velocity $v_{flow}= R_{sh}/t $ (right)
at the shock front.
(g)
Swept-up mass $M_s$ in the shell.
In the RHD case, $M_s$ is estimated
within the outer edge of the shock
front without the precurso.
(h)
Surface density $\sigma_s \equiv M_s / 4 \pi R_{sh}^2$ of the shell.
\label{fig.big4-Iapow}}
\end{figure}

The evolution of the flow is similar to the case of core collapse supernovae.
The Nickel bubble shell has a density
increase toward the contact discontinuity,
with the density contrast reaching $\chi \sim 100$.
The diffusion effect speeds up the freeze-out of the ejecta structure.
A notable difference in the shell is that it is thinner, with a very
prominent precurso 
(Fig.~\ref{fig.prof-Iapow}).
Compared to core collapse supernovae,
the ejecta has a higher ratio of background energy 
to mass ($\propto E/M$),
thus the acceleration of the bubble is weaker.
The expansion rate $a$ rises to a maximum 1.04 before radiation relaxes,
barely faster than the free expansion.
The thickness of the shell reaches  $\beta \sim 0.004$ 
at the turnover. 
 The freezing occurs at a velocity $v_f \sim 7000 \kms$ and 
 a time $t_f \sim 10^{7} \rm \ s$.
 The gas becomes optically thin at $\sim 10^6$ sec.
  By this time, the RHD results
  do not diverge substantially from the HD results,
  but the velocity and the accumulated mass curves had
  started to flatten out,
  indicating that the radiation effect 
  becomes dynamically less important. 

  %

We note that the swept-up mass in the RHD case may be underestimated due to the
large precurso;
 we expect $M_s \sim 0.5\Msun$ as in the HD case.   
For a less amount of $\Ni$, we found that $M_{s} < 0.3\Msun$ 
and $v_{f} \sim 5500 \rm \kms$ for $M_{Ni} = 0.225\Msun$, 
and $M_{s} < 0.2\Msun$ and 
$v_{f} \la 4000 \rm \kms$ for $M_{Ni} = 0.075\Msun$ (Fig.~\ref{fig.mni}).

\subsection{Variation of the Initial Parameters}

\begin{figure}[!hbtp]
\centerline{\includegraphics[width=10cm]{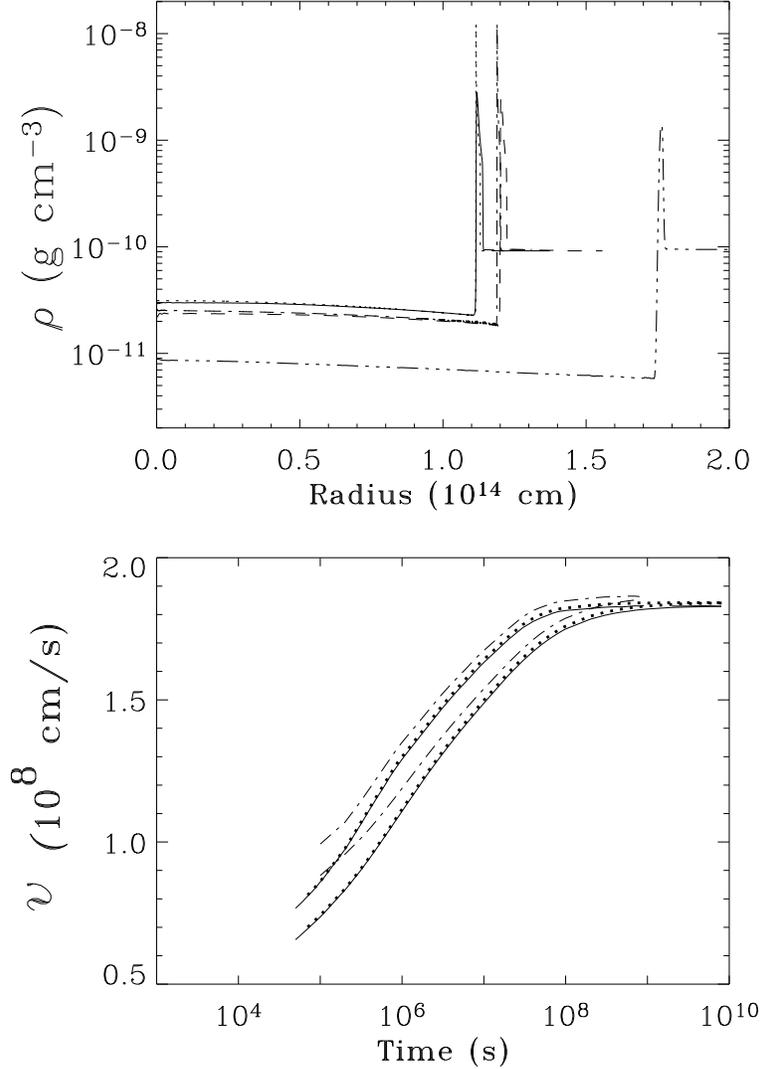}} 
\caption{\small
Density profiles at $10^{6}$ sec (top) and evolution of shell
and flow velocities (bottom) in HD models for core collapse SNe
with varying initial parameters.
The runs in dotted and dash dot lines use a resolution of 8000 zones,
others use a resolution of 2000 zones.
 Solid line:  $t_{0}=10^{4}$ sec, $e_0=2.7\times 10^{11}$ (cgs
               units), and $\omega=3$.
 Dotted line: $t_{0}=10^{2}$ sec, $e_0=2.7\times 10^{18}$ (cgs units), and
              $\omega=3$.
 Dashed line: $t_{0}=10^{4}$ sec, $e_0=2.7\times 10^{11}$ (cgs units), and
              $\omega=1$.
 Dash dot:    $t_{0}=10^{2}$ sec, $e_0=2.7\times 10^{18}$ (cgs units), and
              $\omega=1$.
 Dash dot dot dot: $t_{0}=10^{2}$ sec, $e_0=2.7\times 10^{18}$ (cgs units),
and
                   $\omega=0.1$.
\label{fig.varpar}}
\end{figure}

We carried out simulations with a variety of initial conditions.
We found the solution in the saturated regime is insensitive to
the initial background pressure in the ejecta   
(Fig.~\ref{fig.varpar}). 
In the case of a higher initial background pressure 
 that exceeds
 the accumulated radioactive pressure,
 the sharp structure in the shell takes more time to build up  
 (for which grid resolution also has an effect),
but the solution gradually converges once the radioactive energy input
becomes dominant. 
E.g., the case using
a smaller background pressure $P_{0} = 2.7 \times 10^{10}\dyc$
is no different from the standard case using
$P_{0} = 2.7 \times 10^{18}\dyc$
 (for the core collapse SN model with $t_0=10^2$ sec),
except that the former shows a more pronounced postshock
oscillation trailing in the bubble. 
  Likewise, the solution in our Type Ia model appears insensitive to 
  the variation of background pressure.
 Our case using the standard $\kappa$ shows little differences from 
 the case in Fig. 7. We expect that the insensitivity would sustain
 with a higher value of $\kappa$.
Simulations initialized before $10^{4}$ sec allow the
flow to saturate before the age of interest, and so 
 the results are also insensitive to the initial
 distribution of energy between radiation and material, 
 absorption rate, and details of radiative
 transport.

 We also varied the initial $\Ni$ density contrast.
  We assumed $\Ni$ initially in pressure equilibrium
  with the ambient ejecta, 
  so that its initial background energy
  increases with decreasing $\omega$.
 Thus for $\omega=0.1$ the shell expands to an outer radius.
However, the change from    
$\omega=3$ to $\omega=1$ does not alter 
the shell properties we determined (Fig.~\ref{fig.varpar}).


\section{ANALOGY TO PULSAR WIND NEBULAE}

Chevalier (1984) has considered
a constant-luminosity pulsar bubble interacting with  constant density, freely-expanding supernova ejecta.
We compare his self-similar solution with our hydrodynamical case, 
since the time evolution of the driving power is the only difference.
The self-similarity exists for
a power-law ejecta density $\rho \propto v^{-n}t^{-3}$ and
a time-varying pulsar luminosity $L \propto t^{-l}$, 
when $n \la 3$ and $l \la 1$ (Jun 1998);
the expansion law is 
$r \propto t^{(6-n-l)/(5-n)}$, as obtained from dimensional analysis.
For a constant luminosity $l=0$ and flat ejecta $n=0$, 
$r \propto t^{1.2}$.
The shock cannot be slowed down in the ejecta more
than the free expansion rate, so
the self-similar solutions only exist for $a \ga 1$, i.e. $l \la 1$.

\begin{figure}[!hbtp]
\centerline{\includegraphics[width=10cm]{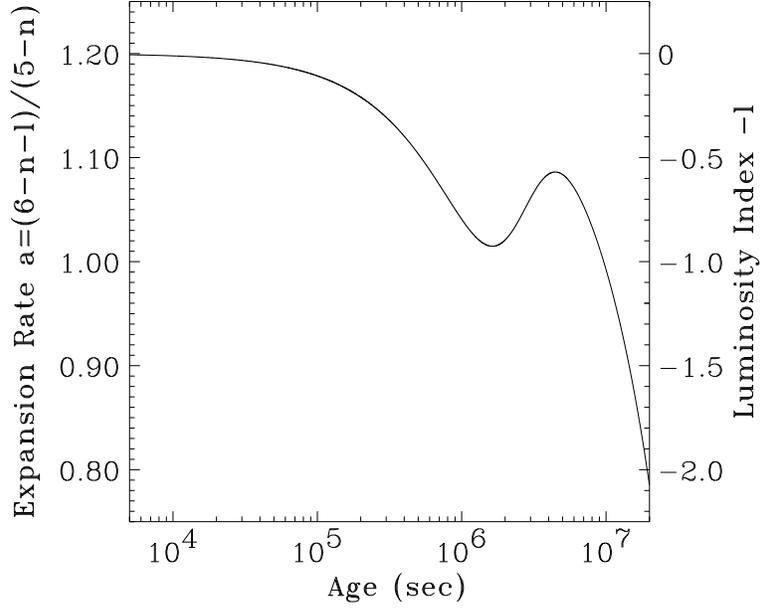}} 
\caption{\small
Luminosity index $l$ of Nickel radioactivity and expansion rate $a$
derived from the self-similar solution, $a=(6-n-l)/(5-n)$.
\label{fig.bradiolumacc}}
\end{figure}

In our solutions, the shell starts to accelerate
 with $a \ga 1$. 
The evolution of the density profile  can thus be compared to 
the self-similar solutions with varying $l$.
In the self-similar solutions, decreasing $l$ (i.e. increasing power input),
 or increasing the expansion rate,
increases the shell thickness. For $l=0$ and $a=1.2$,
we have $\beta = 0.02$.
The shell thickness in our simulations shows a more complicated evolution than 
indicated by the measured expansion rate;
the maximum expansion rate measured occurs at $6\times 10^5$ sec, while
the maximum thickness occurs at $5 \times 10^6$ sec.
The first turnover in the thickness evolution seems to correlate with the 
change in the luminosity power index 
(see Fig.~\ref{fig.bradiolumacc}) at $5 \times 10^6$ sec;
the second turnover occurs after $10^{7}$ sec when the self-similarity breaks down because $l>1$.

\begin{figure}[!hbtp]
\centerline{\includegraphics[width=10cm]{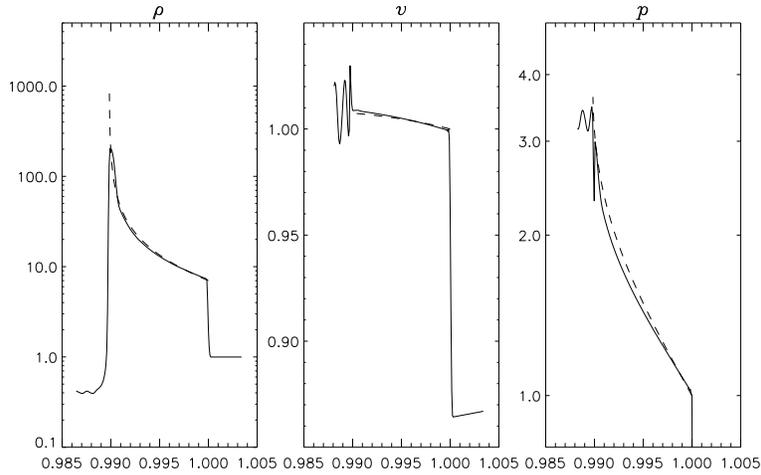}} 
\caption{\small
Hydrodynamical solutions of the Nickel bubble shell in the core collapse SN
model
at $4.0 \times 10^{5}$ sec
overplotted on the self-similar solutions of $\gamma=4/3$, $n=0$, $l=0.3$,
and $a=1.14$. The velocity and pressure show large post shock oscillations
behind the contact discontinuity.
\label{fig.ss-4e5}}
\end{figure}

\begin{figure}[!hbtp]
\centerline{\includegraphics[width=10cm]{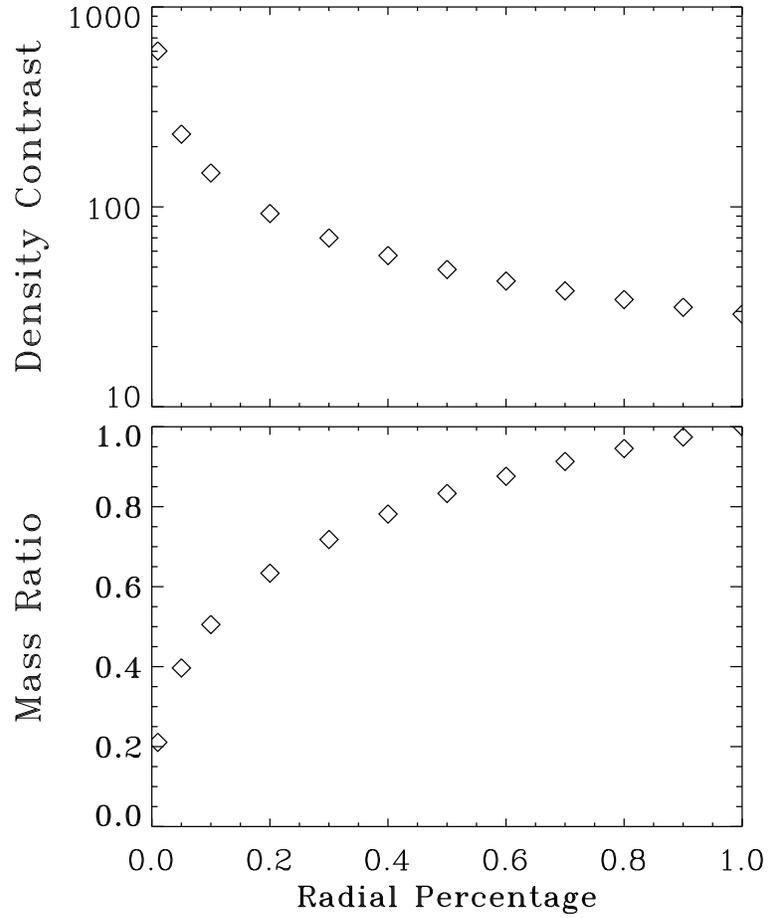}} 
\caption{\small
Top: Density contrast of the self-similar solution at $4 \times 10^{5}$ sec
integrated over a radial fraction of
the shell from the contact discontinuity.
The innermost 1\% region of the shell has an average density contrast over 600,
while the total average is 20.
Bottom: Integrated mass distribution.
\label{fig.ss4e5avg}}
\end{figure}

We overplot our solution at $4 \times 10^5$ sec on the self-similar solution with $l=0.03$ in 
Fig.~\ref{fig.ss-4e5}.
The density distribution in the shell shows a sharp inward increase towards the contact discontinuity,
with over 80\% of the shell mass concentrated within the inner 50\% region in radius 
(Fig.~\ref{fig.ss4e5avg}). 
 The self-similar density is infinite at the contact discontinuity.
The coarseness of the grid obviously sets limits on
the highest density computed in the grid domain.
Our results show a higher density contrast on fixed grids,
because the numerical noise is smaller than on expanding grids.
The shell can be distorted by instabilities (Basko 1994; Jun 1998),
but we expect the density contrast to remain comparable.

Basko's (1994, his Fig. 1) HD solutions  suggest that
the density contrast of the shell is only $\chi \sim 5$ at $4\times 10^5$ sec.
We reproduced his case and found that
the shell broadens to $\beta > 0.1$ at $\sim 4 \times 10^{4}$ s, 
 the turnover is much earlier than in our calculation using
$t_0 = 10^{2}$ s.
In his case using $t_0 = 10^{4}$ s,
the radioactive energy accumulated to $10^4$ s was deposited 
 all at one time in the beginning of the calculation,  
so an excessive acceleration and a
large broadening of the shell were the result.
Basko's Fig. 6  
for the shell surface density   
has the same earlier turnover, 
suggesting this heating effect.

We note that the shell characteristics such as the density contrast 
and the thickness ratio
should only be weakly dependent on the initial $\Ni$ mass, 
since these properties can be approximated
by the self-similar solutions.

The deformation of  the 
bubble-shell interface due to the Rayleigh-Taylor instability
is likely to have a morphology
 similar to that of the pulsar bubble shell studied by Jun (1998)
 in two-dimensional simulations.
 The main difference is that magnetic fields constrain the instability
 in the pulsar bubble case, but are unlikely to be important for our case.
 The Rayleigh-Taylor spikes are directed inward, 
 because of the shell acceleration.
However, 
the instability should eventually end when 
the shell stops accelerating in the comoving stage.
The effect of radiation diffusion is to smooth the pressure gradients, 
which ends the shell acceleration and the instability growth.
The Jun (1998) simulations indicate that the instability does not cause 
radioactive material to have a significantly higher final velocity
(see also Basko 1994).
 
\section{NICKEL BUBBLE SHELL AS EJECTA CLUMPS}

\begin{figure}[!hbtp]
\centerline{\includegraphics[width=10cm]{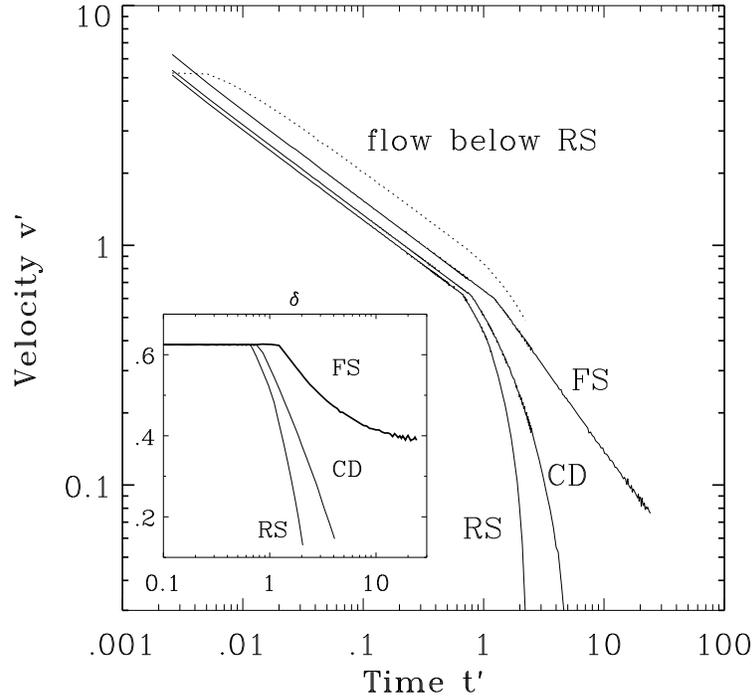}} 
\caption{\small
Evolution of velocities in a supernova remnant with
n=8 ejecta and a constant density ambient medium as described in WC02.
Dotted line:  flow velocity of the ejecta immediately below
the reverse shock (RS). Solid lines: the pattern velocities and deceleration
parameters
$\delta=dlnr/dlnt$ (inset) of the forward shock (FS),
the contact discontinuity (CD), and the reverse shock with time.
The quantities $v'$ and $t'$ are normalized to scaling parameters
given in WC 02.
In our model for core collapse supernovae,
$v = 3162  \ v' \rm \ (km/s)$ and $ t = 1271 \ t' \rm \ (yr)$;
for Type Ia supernovae,
$v = 8452 \  v' \rm \ (km/s)$ and $ t = 244 \ t'  \rm \ (yr)$.
\label{fig.veldecel}}
\end{figure}

We now compare the properties of the Ni bubble shell with the initial
properties of the clumps/bullets inferred from our previous clump-remnant 
simulations for
 Tycho's remnant and the Vela supernova remnant.
 We consider that
the ejecta clumps originate as components from the breakup of the shell,
and examine if the inferred clumps (shell fragments) are able to survive
the crushing in the remnant. 
 The robustness of the clumps can be determined by three parameters:
  the initial density contrast, $\chi$, between the clump and supernova ejecta;
  the initial impact time with the reverse
  shock, related to the initial ejection velocity; and the initial size 
  of the clump.
 The high density contrast created across the shell, $\chi \sim 100$, 
 is compatible with that of ejecta clumps needed
 to survive crushing in the remnant.
In our model for the Vela remnant 
  ($M=10\Msun$, $E_{51}=1$, and $M_{Ni}=0.075\Msun$),
the shell's frozen-in velocity, $v_f \sim 1500 \kms$, is within
the range $ 1000 \la v \la 3000-4000 \kms$ we determined 
as the most likely
origin for the Vela bullets (WC02).
   If the shell has not been greatly disturbed,
   the clumps should be 
   present in bands at the restricted velocity space;
   they are expected to enter the remnant's intershock region 
   at a normalized time $t' \sim 2.2$,
   or $\sim 2800$ yrs 
   (see Fig. 1 in WC02 and Fig.~\ref{fig.veldecel}).
  The velocity-restricted nature may have been detected in  
  the wavelike clumpy structure immediately 
  below Tycho's blast wave front (Reynoso et al. 1999, WC01).
 If the clumps attain a size about 
the thickness of the shell, $\beta=0.04$, 
then at the time of the initial reverse shock impact,
the clumps have an initial size $a_{0} \sim 8\%$ 
in terms of the intershock width,
 or $r_{0} \la 3\%$ in terms of the forward shock radius
(Fig. 1 in WC02).
  Compared to the late clump-remnant interaction in WC02
  that used a larger clump size $a_{0} \sim 1/3$,
  the inferred clumps
  do not appear to have a sufficient strength
  to cause a protrusion on the forward shock.
  Nevertheless, 
  since the computed highest density is limited by numerical resolution,
  a larger density contrast should be achieved in a small region.  
  An increase in the density contrast, e.g., from 100 to 1000,   
  can significantly improve the strength, 
  considering that the cloud crushing time scale is proportional to
 $t_{cc} \propto r_{c}^{-1/2}$ for a fixed mass (WC02, Klein et al. 1994).
  Such small-sized but dense clumps are likely to form, 
  given the steep density profile of the shell.


 In our model for Type Ia supernovae  
($M=1.4\Msun$, $E_{51}=1$, and $M_{Ni}=0.5\Msun$),
 the frozen-in velocity reaches $v_f \sim 7000 \kms$, 
 resulting in a clump-remnant interaction
 initiated in an early state, $t' \sim 1.02$. 
 Compared to the present state $t'\sim 1.7$ of Tycho's remnant,
 the ejection has a strength as indicated for 
 Tycho's knots in our previous 2-D HD simulations (WC01). 
  In addition, the swept-up mass $\sim 0.5 \Msun$ in the shell
  is reasonably in excess of the mass
  0.002 $\Msun$ for the Si+S clump and 0.0004 $\Msun$ for the Fe
  clump in Tycho's SE outline (Hwang, Hughes, \& Petre 1998).            
 However,  there is evidence for Tycho  
that it is  
subluminous (van den Bergh 1993; Schaefer 1996), 
implying a deficiency in $\Ni$.
Observations of SN 1006 suggest an initial Nickel
mass in the range of $0.075-0.16 \Msun$ (Hamilton et al. 1997,
Mazzali et al. 1997).                     
For Sn 1991 bg, Mazzali et al. (1997) estimate
$0.07 \Msun$ of $\Ni$ extending to a velocity of 5,000 $\kms$. 
 In our model, 
 the velocity only extends to $\la 4,000 \ \kms$ with $M_{Ni} = 0.07 \Msun$. 

For core collapse supernovae,
we consider that the object 
 RX J0852--4622 superposed the boundary of the Vela remnant
 is a result of the clump-remnant interaction. 
 WC02 estimated a mass of $0.1 \beta \Msun$, where $\beta < 1$, 
 for the responsible clump,
  which gives a fraction $\la 10\%$ 
  in terms of the total swept-up mass in the shell. 
 This mass fraction appears somewhat large compared to 
  the Si+S clump in Tycho's remnant.
It is unclear how such a massive clump is created, 
if the initial $\Ni$ is not overabundant.

  In our scenario, the initial $\Ni$ has a symmetric distribution
  in the center of supernova ejecta. 
 It should be noted that 
  dynamical processes prior to the Nickel bubble effect,
 e.g., neutrino-driven convection (Kifonidis et al. 2000), 
 may give rise to an anisotropic distribution of elements. 
   If $\Ni$ is mixed  outwards during the explosion,
   the amount of radioactive
   material inside each individual bubble could be orders of magnitude
   lower;                                     
 the subsequent radioactive heating may cause merging between Nickel bubbles, 
 and so give further compression. 
 This process remains to be
 examined by future multidimensional calculations. 
 Our result here should be noted as a 
 lower limit on the ejecta-clump properties for the clump-remnant interaction. 

\section{CONCLUSIONS}\label{sec:conclu}

We investigated the ejecta structure in supernovae due to 
the radioactive heating
 from the $\Ni \rightarrow\ \co \rightarrow\ \fe$ decay sequence, 
under the assumption of spherically symmetric flow.
Two approaches were used: hydrodyamics (HD) simulations and 
radiation hydrodynamics (RHD) simulations. 
The RHD calculation included a proper treatment of radiation transport 
prior to the age $10^7$ sec, when 
about 25\% of the total radioactive energy is still to be released. 
Radiative effect becomes 
dynamically less important as
the evolution tends towards the optically thin regime.
 Based on the mutual coherence in the HD and RHD calculations,
 we obtained the properties of the shell driven by the radioative
 expansion.
 We believe that going to later stages of diffusion would not change 
 this conclusion.


 The expansion of the Nickel bubble
 can sweep up a dense shell of 
 $\la 1 \Msun$ shocked ambient ejecta in
 core collapse and 
 Type Ia supernovae, 
 with the highest density in the shell 
 over 100 times the ambient ejecta density.
The shell has an inward density increase toward the bubble-shell interface, 
and the highest density
contrast computed is limited by numerical resolution. 
 For the hydrodynamical solution,
 the structure of the shell can be approximately described
 by a self-similar solution.
  For the RHD solution, the main difference is that radiative diffusion 
  gives rise to 
  a broader and less dense shell that freezes out in the ejecta.
   Because the radioactive pressure eventually dominates, 
   the saturated solution is not sensitive to
   the initial conditions such as the background pressure and 
   the initial density contrast of Ni; 
  the only crucial factor to
  the solution is the initial $\Ni$ abundance.                         
  We presume that a higher background pressure due to the uncertainty of the
  adiabat would not affect the results.

  We examined if the properties of the shell fragments are comparable to that of
  the ejecta clumps protruding the outlines of Tycho's remnant 
  and the Vela remnant. 
  The density contrast created across the shell, $\chi \sim 100$,  
  is compatible with that of the ejecta clumps needed to survive the
  clump-remnant interaction.
  In our standard Type Ia model, 
  the shell's frozen-in velocity attained at $\sim 10^6$ sec appears
  sufficient to give an ejection as indicated for Tycho's knots,
  in spite that most ($\la 75 \%$)
  of the radioactive energy is still to be released. 
  For core collapse supernovae like the Vela remnant, 
  however, 
  the small size and the late ejection  
  of the inferred clump result in a weak strength to resist 
 crushing and further cause protrusions on the remnant outline. 
%
%
%

I am grateful to Roger Chevalier for supporting this work, 
and Toshi Shigeyama for useful
correspondence on Type Ia supernovae.
I thank the referee for careful and constructive comments on the manuscript.
The computations were carried out on the IBM SP2 at University
of Virginia, and the PC cluster 
provided by Taiwan's NSC
grant NSC93-2112-M033-002. Support for this work was also provided by 
NSC grant NSC93-2112-M033-013.. 


\clearpage

\vspace{10mm}

\vspace{10mm}

\vspace{10mm}

\vspace{10mm}

\vspace{10mm}

\vspace{10mm}

\vspace{10mm}

\vspace{10mm}

\vspace{10mm}

\vspace{10mm}

\vspace{10mm}

\vspace{10mm}


\clearpage

\begin{thebibliography} {}


 \bibitem[]{}
 Audi, G., Wapstra, A. H., \& Thibault, C. 2003, Nucl. Phys. A729, 337
 (http://ie.lbl.gov/toimass.html for the atomic mass)

 \bibitem[]{}
 Aschenbach, B., Egger, R., \& Tr\"umper, J. 1995, Nature, 373, 587

 \bibitem[]{}
 Basko, M. 1994, \apj, 425, 264


 \bibitem[]{}
 Chevalier, R. A. 1976, \apj, 207, 872 


 \bibitem[]{}
 Chevalier, R. A. 1984, \apj, 280, 797




 \bibitem[]{}
 Chevalier, R. A., \& Liang, E. P. 1989, \apj, 344, 332




 \bibitem[]{}
 Dwarkadas, V., \& Chevalier, R. A. 1998, \apj, 497, 810

\bibitem[]{}
Fireston, R. B., \& Shirley, V. S., ed 1996, Table of Isotopes (8th ed.; John Wiley \& Sons, Inc.) 
 
\bibitem[]{}
Lide, D. R., ed 1992, CRC Handbook of Chemistry and Physics (73d ed.; Boca Ration: CRC Press), 11-38

 \bibitem[]{}
 Filippenko, A. V., \& Sargent, W. L. W. 1989, \apj, 345, L43


 \bibitem[]{}
 Hamilton, A. J. S., Fesen, R. A., Wu, C.-C., Crenshaw, D. M.,
 \& Sarazin, C. L. 1997, ApJ, 482, 838




 \bibitem[]{}
 Hwang, U., Hughes, J. P., \& Petre, R. 1998, \apj, 497,833

\bibitem[]{}
Jun, B.-I. 1998, \apj, 499, 282 



  \bibitem[]{}
  Kifonidis, K., Plewa T., Janka, H.-Th, \& M\"uller, E. 2000, \apj, 531, L123

 \bibitem[]{}
 Klein, R. I., McKee, C. F., \& Colella, P. 1994, \apj, 420, 213


\bibitem[]{}
Li, H., McCray, R., \& Sunyaev, R. A. 1993, \apj, 419, 824

\bibitem[]{}
Lide, D. R., ed 1992, CRC Handbook of Chemistry and Physics (73d ed.; Boca Ration: CRC Press), 11-38
 
 \bibitem[]{}
 Matheson, T., Filippenko, A. V., Ho, L. C., Barth, A. J., \& Leonard, D. C.
 2000, \aj, 120, 1499



 \bibitem[]{}
 Mazzali, P. A., Chugai, N., Turatto, M., Lucy, L. B., Danziger, I. J.,
 Cappellaro, E., Della Valle M., \& Benetti, S. 1997, MNRAS, 284, 151

 \bibitem[]{}
 McCray, 1993 \apj, 510, 379

 \bibitem[]{}
 Nomoto, K., Thielemann, F.-K.,  \& Yokoi, K. 1984, ApJ, 286, 644

\bibitem[]{}
Reynoso, E. M., Vel'azquez, P. F., Dubner, G. M., \& Goss, W. M. 1999, AJ, 117, 1827.

\bibitem[]{}
Schaefer, B. E. 1996, ApJ, 459, 438

\bibitem[]{}
Shigeyama, T., \& Nomoto, K. 1990, \apj, 360, 242

 \bibitem[]{}
 Spyromilio, J. 1994, MNRAS, 266, L61


\bibitem[]{}
Stathakis, R. A., Dopita, M. A., Cannon, R. D., \& Sadler, E. M.
1991, in Supernovae, ed. S. E. Woosley (New York: Springer) 95


\bibitem[]{}
Stone, J. M., Mihalas, D., \& Norman, M. L. 1992, ApJS, 80, 819

\bibitem[]{}
van den Bergh, S. 1993, ApJ, 413, 67


\bibitem[]{}
Wang, C.-Y., \& Chevalier, R. A. 2001, \apj, 549, 1119 (WC01)

\bibitem[]{}
Wang, C.-Y., \& Chevalier, R. A. 2002, \apj, 574, 155 (WC02)


 \bibitem[]{}
 Winkler, P. F., Tuttle, J. H.,  Kirshner, R. P., \& Irwin, M. J.
  1988, in Supernova Remnants and the Interstellar Medium, ed. R. S.
  Roger \& T. L. Landecker (Cambridge: Cambridge Univ. Press), 65

\bibitem[]{}
Woosley, S. E. 1988, \apj, 330, 218

\bibitem[]{}
Woosley, S. E., Pinto, P. A., \& Hartmann, D. 1989, \apj, 346, 395 

\end{thebibliography}
\end{document}